\documentclass[12pt,preprint]{aastex}

\newcommand{\lta}{{\>\rlap{\raise2pt\hbox{$<$}}\lower3pt\hbox{$\sim$}\>}}
\newcommand{\gta}{{\>\rlap{\raise2pt\hbox{$>$}}\lower3pt\hbox{$\sim$}\>}}
\shorttitle{Detecting Reionization Sources}
\shortauthors{Stiavelli et al.}

\begin{document}

%% LaTeX will automatically break titles if they run longer than
%% one line. However, you may use \\ to force a line break if
%% you desire.

\title{Possible Detection of Cosmological Reionization Sources}

%% Use \author, \affil, and the \and command to format
%% author and affiliation information.
%% Note that \email has replaced the old \authoremail command
%% from AASTeX v4.0. You can use \email to mark an email address
%% anywhere in the paper, not just in the front matter.
%% As in the title, you can use \\ to force line breaks.

\author{M. Stiavelli, S. Michael Fall,  and N. Panagia\altaffilmark{1}}
\affil{Space Telescope Science Institute, 3700 San Martin Drive, Baltimore, MD
21218}

%% Notice that each of these authors has alternate affiliations, which
%% are identified by the \altaffilmark after each name.  Specify alternate
%% affiliation information with \altaffiltext, with one command per each
%% affiliation.

\altaffiltext{1}{ESA Space Telescope Division}

%% Mark off your abstract in the ``abstract'' environment. In the manuscript
%% style, abstract will output a Received/Accepted line after the
%% title and affiliation information. No date will appear since the author
%% does not have this information. The dates will be filled in by the
%% editorial office after submission.

\begin{abstract}

We compare the available catalogs of $z\approx6$ galaxies in the
Hubble Ultra-Deep Field (UDF) and in the Great Observatories Origins
Deep Survey (GOODS) with the expected properties of the sources of
cosmological reionization from our previous theoretical study. Our
approach is based on the mean surface brightness of the sources
required for reionization and depends on relatively few undetermined
parameters. We find that the observed mean surface brightness of
galaxies at $z \approx 6$ is sufficient for reionization, provided
that the sources are composed of hot metal-free or metal-poor stars,
regardless of whether reionization occurs over a short or long
interval of redshift.  The broad agreement between the new
observations and our predictions suggests that we may have detected
the sources responsible for some or even all of the reionization of
hydrogen.

\end{abstract}

%% Keywords should appear after the \end{abstract} command. The uncommented
%% example has been keyed in ApJ style. See the instructions to authors
%% for the journal to which you are submitting your paper to determine
%% what keyword punctuation is appropriate.


%% From the front matter, we move on to the body of the paper.
%% In the first two sections, notice the use of the natbib \citep
%% and \citet commands to identify citations.  The citations are
%% tied to the reference list via symbolic KEYs. The KEY corresponds
%% to the KEY in the \bibitem in the reference list below. We have
%% chosen the first three characters of the first author's name plus
%% the last two numeral of the year of publication as our KEY for
%% each reference.

\section{Introduction}

   The reionization of the intergalactic medium (IGM) was undoubtedly
one of the most significant events in cosmic history. It completely
changed the opacity of the universe to ionizing radiation and may also
have influenced the formation and evolution of galaxies and other
structures (see, e.g., Loeb \& Barkana 2002).  The absence of strong
Gunn-Peterson Ly$\alpha$ absorption in the spectra of distant quasars
indicates that reionization was complete by $z \approx 6$
\citep{becker01, fan02, white03}, while the polarization of the cosmic
microwave background (CMB) radiation indicates that it began at higher
redshifts \citep{spergel03}. A major goal of extragalactic astronomy
is to detect and characterize the sources of UV radiation responsible
for reionization.

    In a previous paper, we outlined a method to guide and interpret
searches for the reionization sources with existing and planned
telescopes (Stiavelli, Fall, \& Panagia 2004, hereafter Paper I).  Our
work builds on several important papers in this field, including those
by \citet{miraldarees} and Madau, Haardt, \& Rees (1999).  We consider
the same physical processes as previous authors, but we focus on
low-metallicity sources (Population II and III stars), which are more
efficient ionizers and are perhaps more natural at high redshifts
(e.g., Malhotra \& Rhoads 2002).  We ignore active galactic nuclei
(AGN) because constraints from the X-ray background indicate that they
make a relatively small contribution to reionization (Dijkstra et
al. 2004).  We express our results in terms of the mean surface
brightness of the reionization sources instead of their volume
emissivity or star formation density because this allows a more direct
comparison with observations.  We focus on the expected location of
the sources in a plot of mean surface number density against apparent
magnitude. Only a moderately narrow band in this diagram is allowed by
the demands of producing enough ionizing radiation but not too many
heavy elements (for stellar sources).

   We showed that the existing deep surveys with the {\it Hubble Space
Telescope} ({\it HST}), such as the Hubble Deep Fields (HDFs: Williams et
al. 2000) and the Great Observatories Origins Deep Survey (GOODS:
Giavalisco et al. 2004), do not probe far into the allowed part of
this diagram, while future deep surveys with the {\it James Webb Space
Telescope} ({\it JWST}) will do so easily. We also pointed out that there was
a reasonable chance of detecting the reionization sources with the
then-planned {\it HST} Ultra-Deep Field (UDF: Beckwith et al., in
preparation). This observing program has now been completed, the data
have been released, and several papers have appeared with catalogs of
sources \citep{bouwens04,bunker04}. Here, we combine the available UDF
and GOODS observations with our previous theoretical expectations to
search for the reionization sources.  Throughout this paper, we adopt
values of the cosmological parameter derived from the {\it Wilkinson
Microwave Anisotropy Probe} ({\it WMAP}): $\Omega_\Lambda = 0.732$,
$\Omega_m = 0.268$, $\Omega_b = 0.044$, and $H_0 = 71$ km s$^{-1}$
Mpc$^{-1}$ \citep{spergel03}

\section{Models}

Our general method is applicable to a wide variety of reionization
sources. For specific predictions, however, we focus on sources
composed either of metal-free stars (hereafter Population~III) or
metal-poor stars ($10^{-3} \lta Z/Z_\odot \lta 10^{-2}$, hereafter
Population~II). We approximate the spectral energy distributions of
these sources by blackbodies with temperatures of $10^5$~K for
Population~III and $5\times 10^4$~K for Population~II. These
effective temperatures are most appropriate for stellar populations
with top-heavy initial mass functions (IMFs dominated by stars more
massive than $\sim 30 M_\odot$). Since Population~III stars are hotter
than Population~II stars, they produce more ionizing photons for a
given flux at longer wavelengths.  In this sense, Population~III stars
are also more efficient ionizers than AGNs.

Figure~1 shows the expected cumulative mean surface number density of
reionization sources as a function of their apparent AB magnitude in
the non-ionizing UV continuum at a rest-frame wavelength of
1400~\AA. Here we have assumed that the comoving volume density of the
sources is constant over the range of redshifts $5.8 \lta z \lta 6.7$
spanned by $i$-band dropouts in the UDF and GOODS (see below). The
luminosity function of the sources is assumed to have the Schechter
form, parameterized by its knee $M_*$ and slope $\alpha$. For
reference, the Lyman-break galaxies at $z = 3$ have $M_{*,1400} =
-21.2$ and $\alpha = 1.6$. \citep{steidel99, yan02}. These are the
parameters adopted for the top panels of Figure~1. The middle panels
have a brighter knee ($M_{*,1400} = -23.2$), and the bottom panels
have a steeper slope ($\alpha = 1.9$). The predictions in the left
panels are for Population~III stars; those in the right panels are for
Population~II stars. (Figure~1 here is a rearrangement of several
panels in Figures~2 and 3 of Paper I.)

The requirement that the sources be able to ionize all the hydrogen 
in the IGM corresponds, for a given spectral shape, to a definite 
mean surface brightness at any chosen wavelength ($\lambda_{\rm rest} 
= 1400$~\AA\ in all cases considered here). This surface brightness
depends on the fraction $f_c$ of Lyman-continuum photons that escape 
from the sources and on the clumpiness $C$ of the IGM, which in turn 
determines the recombination rate. The surface brightness required
for reionization fixes the normalization of the curves in Figure~1,
which are labeled by the corresponding parameters $(f_c,C)$. We 
have calculated these curves from the equation of ionization 
balance, including both H and He, and the effect of recombinations 
(as described in detail in Paper I).

The minimum mean surface brightness of the reionization sources
is given by $(f_c,C) = (1,1)$; in this case, all Lyman-continuum
photons escape and the recombination rate is minimum. For stellar 
sources, there is also a maximum mean surface brightness,
given by the condition that they not produce too many heavy elements.
We adopt the generous but non-rigorous limit $Z \lta 0.01 Z_{\odot}$ on
the cosmic mean metallicity (the mean density of heavy elements in
stars and interstellar and intergalactic matter divided by the mean
baryon density) at $z = 6$ (explained in detail in Paper I). 
The shaded regions in Figure~1 are excluded by
these constraints, while the clear regions are allowed.

The specific predictions shown in Figure~1 assume that reionization is
complete at $z=6$. This is supported strongly by the Ly$\alpha$
absorption of $z\sim6$ quasars and by a variety of theoretical
arguments and hydrodynamical simulations
\citep{haimanholder03,gnedin04}.  
The mean surface brightness required
for reionization also depends on the interval of redshift over which
the sources are active. Here we concentrate on the interval $\Delta
z\approx1$ just above $z\approx6$.  This closely matches the interval
spanned by $i$-band dropouts and is marginally compatible (at
2$\sigma$ confidence) with the measured CMB polarization
\citep{spergel03}.  The interval $\Delta z \approx 1$ is also likely a
worst-case (i.e., minimum-surface-brightness) scenario in the sense
that there could be additional reionization sources at redshifts above
$z \approx 7$, which would relax the requirements on those
below. However, the dependence of the mean surface brightness required
for reionization on the redshift interval is relatively weak,
increasing by only 0.1, 0.4, and 0.8 magnitudes as $\Delta z$
increases from 1 to 3, 1 to 10, and 1 to 30 (for $C\approx 1$).

\section{Observations}

We now compare the predictions of Paper I, summarized in the previous
section, with recent observations from the UDF and GOODS.  The UDF
observations were made in 400 orbits on a single pointing with the
wide-field channel (WFC) of the Advanced Camera for Surveys (ACS).
The two longest integrations, 144 orbits each, were made in the
$i$-band (F775W filter) and $z$-band (F850LP filter).  We base our
analysis on the $i$-band dropouts ($z \approx 6$ galaxies) identified
by Bunker et al. (2004) using the Sextractor program \citep{bertin} in
the final combined images.  The magnitudes and colors of the objects
were measured in apertures 0.5 arcsec in diameter.  The UDF catalog
includes 53 objects with $i-z \geq 1.3$ and 22 with $i-z \geq 2.0$
down to a limiting magnitude of $z=28.5$ at $S/N=8$, where the
incompleteness is only 2\%\ (Table~1 of Bunker et al. 2004).  We
analyze the $i-z \geq 1.3$ and $i-z \geq 2.0$ subsamples separately.
The latter may miss some high-reshift objects, but it should be nearly
free of contamination by low-redshift objects.  We have verified that
the same selection criteria produce essentially the same samples of
objects when applied to the $z$-detected catalog released by the UDF
team (Beckwith et al., in preparation; Stiavelli et al., in
preparation).

We combine these UDF observations with wider and shallower, but
otherwise similar, observations from GOODS to obtain more reliable
estimates of the bright part of the surface density-magnitude
relation.  For consistency with the UDF catalog, we again use
magnitudes and colors measured in apertures 0.5 arcsec in diameter,
taken from the V1.0 GOODS catalog (Dickinson et al. 2004; Giavalisco
et al., in preparation).  This includes 77 objects with $i-z \geq 1.3$
and 14 with $i-z \geq 2.0$ down to a limiting magnitude of $z=26.8$ at
$S/N = 10$.  We correct these counts for incompleteness, which
increases from 10\%\ at $z=26.0$ to 40\%\ at $z=26.5$ for compact
sources.  We use the GOODS observations brighter than $z=26.5$ and the
UDF observations fainter than this magnitude.  The expected
uncertainties in the surface density due to Poisson fluctuations alone
range from 60\%\ at $z=24$ to 20\%\ at $z = 28$, while those due to
large-scale structure range from 20\%\ to 40\%\ at the same magnitudes
\citep{somerville,bunker04}.  We do not correct the observed mean
surface brightness for objects fainter than our detection limits in
luminosity and/or surface brightness. In this sense, all our empirical
estimates are lower limits to the true mean surface brightness.

Before comparing these observations with our predictions, we must
examine the relation between the measured $z$-band magnitudes and the
rest-frame 1400~\AA\ magnitudes adopted in Figure~1.  We estimate
the offset between these magnitudes by convolving the total $z$-band
response function with $10^5$~K and $5 \times 10^4$~K black-body
spectra truncated below rest-frame 1216~\AA\ and redshifted through the
interval $5.8 < z < 6.7$. This procedure mimics observations of
reionization sources composed of Population~III and Population~II
stars with heavy HI absorption by some combination of stellar,
interstellar, or intergalactic material. The resulting offsets between
the $z$-band and rest-frame 1400~\AA\ magnitudes are -0.03 mag for
Population~III sources and +0.02 mag for Population~II sources. Such
small offsets are the product of a fortuitous near-cancellation.
Sources at $z\approx 5.8$ appear brighter in the $z$-band (by few
tenths of a magnitude) than at rest-frame 1400~\AA\ because their blue
UV continua are measured at shorter wavelengths, which contributes a
negative correction. However, objects at $z \approx 6.7$ appear
fainter than at rest-frame 1400~\AA\ because the $z$-band samples
mostly the part of the spectrum attenuated by HI, which contributes a
positive correction.  Thus, for a uniform distribution of redshifts
within the interval $5.8 \lta z \lta 6.7$, the offsets from the two
extremes tend to cancel out.  Given these small differences 
between the $z$-band and rest-frame 1400~\AA\ magnitudes, we simply
ignore them in the following.

The resulting cumulative surface density-magnitude relation is shown
by the stepped lines in each panel of Figure~1.  The upper
(light-gray) line refers to the $i-z\geq1.3$ color cut, while the
lower (dark-gray) line refers to the more stringent $i-z\geq2.0$ color
cut. We expect the true relation to lie between the two stepped
lines. Both of these relations appear generally compatible with our
predictions, especially when allowance is made for the large
statistical uncertainties at the bright ends. The observed relation
extends into the shaded region at the bright end, but we have checked
that this does not violate the global metallicity constraint, which
pertains to the mean surface brightness when integrated over all
magnitudes.  We find that the integrated surface brightness of the
objects brighter than our limiting magnitude ($z_{\rm lim} = 28.5$) is
$\mu_{AB}=25.4$ and 26.7 mag arcmin$^{-2}$, respectively, for the
$i-z\geq1.3$ and $i-z\geq2.0$ subsamples. These are consistent with our
predicted minimum surface brightness, $\mu_{AB}=28.8$ and 27.2 mag
arcmin$^{-2}$, for reionization by Population~III and Population~II
sources, respectively, and a redshift interval $\Delta z=1$ (Paper I).
We have repeated the entire analysis above with the objects selected
by the same criteria from the catalog released by the UDF team and
obtained the same values of the mean surface brightness to within 10\%.

\section{Comparisons}

Our conclusion about the ability of the sources detected in the UDF to
reionize the IGM differs from the one reached by Bunker et al. (2004).
The reason for this is that the two studies, which are based on the
same observations, adopt different conditions for reionization; ours
is taken from Paper I, theirs from equation (27) of Madau et
al. (1999).  These conditions involve different assumptions about the
temperature $T_{\rm IGM}$ of the IGM and the ionizing efficiencies of
the sources as given by the ratio $\epsilon_{\rm ion} \equiv \dot
N_{\rm ion}/{(\nu L_\nu)}_{1400}$ of the production rate of ionizing
photons to the luminosity in the non-ionizing UV continuum at
1400~\AA, which depends on the effective temperature and hence the
metallicity and IMF of a stellar population (Baraffe \& El Eid 1991;
Tumlinson \& Shull 2000; Bromm, Kudritzki, \& Loeb 2001; Schaerer
2003).  As in Paper I, we adopt $T_{\rm IGM} = 20,000$~K and
$\epsilon_{\rm ion} = 2 \times 10^{11}$ erg$^{-1}$ (Population III) or
$\epsilon_{\rm ion} = 4.4 \times 10^{10}$ erg$^{-1}$ (Population II),
while Madau et al. (1999) adopt $T_{\rm IGM} = 10,000$~K and
$\epsilon_{\rm ion} \approx 7 \times 10^9$ erg$^{-1}$ (Population I,
with solar metallicity and a Salpeter IMF).  Apart from these differences,
the two conditions are almost identical (within $\sim$10\%), as we
have verified by test calculations with the same values of $T_{\rm
IGM}$, $\epsilon_{\rm ion}$, $f_c$, and $C$.  The ionization
conditions are, however, expressed differently: ours in terms of a
surface brightness, theirs in terms of a volume emissivity or star
formation density.

The temperature of the IGM and the metallicity and IMF of the stellar
population can each change the mean surface brightness required for
reionization by factors of a few. The recombination coefficient and
hence the required ionization rate is a factor of 2 lower for $T_{\rm
IGM} = 20,000$~K than for $T_{\rm IGM} = 10,000$~K. The higher IGM
temperature seems more appropriate for several reasons.  The
equilibrium temperature of metal-poor HII regions excited by hot stars
is $\sim$20,000~K (see, e.g., Osterbrock 1989). This is also close to
the inferred temperature of the IGM at $2 \la z \la 4$ (Zaldarriaga,
Hui, \& Tegmark 2001).  In the simplest non-equilibrium models (those
with little or no late heating), the IGM cools following reionization,
and has $T_{\rm IGM} \ga 20,000$~K at $z \approx 6$
\citep{theuns02,huihaiman}. The ionizing efficiency of a stellar population
increases by a factor of 3 for a Salpeter IMF and a factor of 10 for a
top-heavy IMF as the metallicity decreases from solar to zero
metallicity \citep{schaerer03}.  The mean metallicity, averaged over
all galaxies, is roughly solar in the present-day universe and must
have been lower at the epoch of reionization, although we do not know
by exactly how much.  The ionizing efficiency of a stellar population
also depends on the IMF, increasing by factors of 3 (for solar
metallicity) to 10 (for zero metallicity) between a Salpeter IMF and a
top-heavy IMF \citep{bruzualcharlot03, schaerer03}.  The latter is
favored by recent theoretical studies of primordial star formation
(e.g., Abel, Bryan, \& Norman 2000; Bromm, Coppi, \& Larson 2002).

The shortfall of ionizing photons estimated by Bunker et al. (2004)
using the Madau et al. (1999) condition is at least a factor of 3.  The
different assumptions we have made about the IGM temperature, stellar
metallicity, and IMF each reduce the mean surface brightness required
for reionization by factors of 2, 3-10, and 3-10, respectively. Any
two of these factors would be enough to satisfy the reionization
condition and all three taken together (as we have assumed in Paper I)
do so with a comfortable margin.

\section{Conclusions}

Our main conclusion is that the surface brightness-magnitude relation
for $z \approx 6$ objects detected in the UDF and GOODS agrees
remarkably well with our predicted relation for metal-poor
reionization sources. In particular, the observed integrated surface
brightness is consistent with that required for reionization with
reasonable values of the escape fraction ($0.05<f_c<0.5$) and
clumpiness parameter ($1<C<30$) if the sources are composed either of
Population~III or Population~II stars with a top-heavy IMF.
Reionization by sources with a Salpeter IMF, although possible, is
more difficult, requiring both a high escape fraction and a very low
metallicity.

We find that the shape of the observed surface brightness-magnitude
relation at $z \approx 6$ is roughly similar to that of the
Lyman-break galaxies at $z \approx 3$ (consistent with the conclusions
of Bouwens et al. 2004 and Bunker et al. 2004). However, we obtain
somewhat better agreement if the luminosity function of the
$z\approx6$ galaxies has a brighter knee and a steeper slope.  It is
worth noting that, as the slope approaches the divergence value
$\alpha = 2$, progressively more of the ionizing radiation comes from
fainter, possibly undetected sources.  Population II sources require
higher values of $f_c$ and lower values of $C$ than Population III
sources, because cooler stars are less efficient ionizers. It is not
clear, however, whether sufficiently high values of $f_c$ could be
reached in sources that were already enriched in metals and dust and
therefore likely to have more UV absorption.

The predictions presented here are based on the assumption that
reionization occurs over the relatively small interval of redshift
$\Delta z \approx 1$ just above $z \approx 6$. It is remarkable that,
even in this most demanding case, the observed objects are sufficient
to reionize the IGM. Of course, it is possible that reionization began
much earlier.  The polarization of the CMB measured by {\it WMAP}
indicates that the IGM was at least partly ionized at $z\sim 20$
(albeit with large uncertainties, Spergel et al. 2003).  However, the
high temperature of the IGM at $z \approx 4$ inferred from quasar
absorption-line systems indicates that much of the heating
accompanying reionization occurred at $z << 20$ \citep{theuns02,huihaiman}.
Some theoretical models suggest that reionization might have occurred
more than once, for example at $z \approx 15$ and then again at
$z\approx6$ \citep{cen03}.  We conclude that, even if the $z \approx
6$ objects observed in the UDF and GOODS were not the only sources
responsible for the reionization of the IGM, they would have
contributed substantially to its final stages.

An important future development will be to extend of this analysis 
to observations at infrared wavelengths. In this way, we could attempt
to detect additional reionization sources at redshifts beyond $z\approx7$ 
and to determine their contribution relative to the sources at lower
redshift discussed here.  Such observations in and near the UDF have 
already been made with the Near Infrared Camera and Multi-Object
Spectrometer (NICMOS) on {\it HST} and will become available soon (Thompson 
et al., in preparation). Further progress could be made with the Wide 
Field Camera 3 if and when it is installed on {\it HST}. However, a complete
census of all the reionization sources, especially any at 
$z \gta 12$, must await the launch of {\it JWST}. 

%% If you wish to include an acknowledgments section in your paper,
%% separate it off from the body of the text using the \acknowledgments
%% command.

%% Included in this acknowledgments section are examples of the
%% AASTeX hypertext markup commands. Use \url without the optional [HREF]
%% argument when you want to print the url directly in the text. Otherwise,
%% use either \url or \anchor, with the HREF as the first argument and the
%% text to be printed in the second.

\acknowledgments

We are grateful to the UDF and GOODS teams for obtaining and releasing
the data on which this study is based.  We also thank Daniel Schaerer
for making his stellar population models available to us in electronic
form. {\it HST} data are obtained at STScI, which is operated by AURA,
Inc., under NASA contract NAS5-26555.  This work is partially
supported by NASA Grant NAG5-12458.

%% The reference list follows the main body and any appendices.
%% Use LaTeX's thebibliography environment to mark up your reference list.
%% Note \begin{thebibliography} is followed by an empty set of
%% curly braces.  If you forget this, LaTeX will generate the error
%% "Perhaps a missing \item?".
%%
%% thebibliography produces citations in the text using \bibitem-\cite
%% cross-referencing. Each reference is preceded by a
%% \bibitem command that defines in curly braces the KEY that corresponds
%% to the KEY in the \cite commands (see the first section above).
%% Make sure that you provide a unique KEY for every \bibitem or else the
%% paper will not LaTeX. The square brackets should contain
%% the citation text that LaTeX will insert in
%% place of the \cite commands.

%% We have used macros to produce journal name abbreviations.
%% AASTeX provides a number of these for the more frequently-cited journals.
%% See the Author Guide for a list of them.

%% Note that the style of the \bibitem labels (in []) is slightly
%% different from previous examples.  The natbib system solves a host
%% of citation expression problems, but it is necessary to clearly
%% delimit the year from the author name used in the citation.
%% See the natbib documentation for more details and options.

\clearpage

%% Use the figure environment and \plotone or \plottwo to include 
%% figures and captions in your electronic submission.

\begin{figure}
\plotone{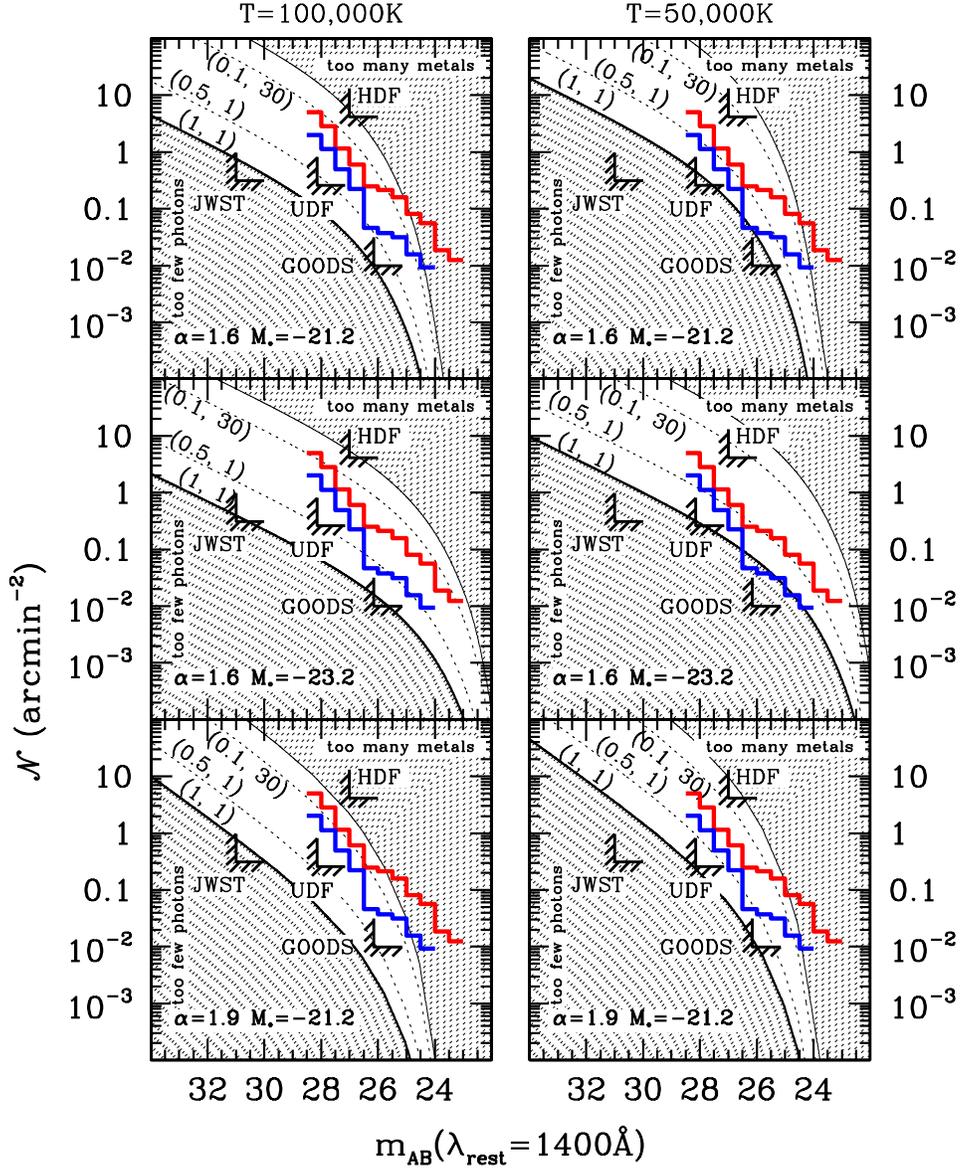}

\caption{Cumulative mean surface number density vs apparent rest-frame
1400\AA\ AB magnitude of reionization sources with different luminosity
functions and effective temperatures. In each panel, the lower solid
line represents the minimum-surface brightness model, with $(f_c,C)=
(1,1)$, while the upper thin solid line represents the global
metallicity constraint.  The thin dotted lines correspond to models
with $(f_c,C)=(0.5,1)$ and (0.1, 30). The thick stepped lines
represent the observations from the UDF and GOODS with the two color
cuts: $i-z\geq1.3$ (light gray) and $i-z\geq2.0$ (dark gray).  The
L-shaped markers delimit the areas probed by GOODS, HDF, UDF, and an
ultra-deep survey with {\it JWST}. The adopted parameters of the
luminosity function are indicated in each panel.  The left and right
panels refer to sources composed, respectively, of Population III and
II stars with a top-heavy IMF. }

\end{figure}

\end{document}